\documentclass[12pt, draftclsnofoot, onecolumn]{IEEEtran}
\usepackage[dvips]{graphicx}
\usepackage{amsfonts,epsfig,cite,array,multirow,graphicx,amsmath,amsthm,ltablex,tabularx,setspace,arydshln,amssymb,multirow}
\begin{document}
\title{Performance of QAM Schemes with Dual-Hop DF Relaying Systems over
Mixed $\eta$-$\mu$ and $\kappa$-$\mu$ Fading Channels}
\author{Dharmendra Dixit  and P. R. Sahu~\IEEEmembership{Member,~IEEE}
\thanks{ Authors are with the
School of Electrical Sciences, Indian Institute of Technology Bhubaneswar, Orissa, India
(e-mail: dd12, prs)@iitbbs.ac.in.} }
\maketitle
\begin{abstract}
Performance of  quadrature amplitude modulation (QAM) schemes is analyzed with dual-hop decode-and-forward (DF) relaying systems over mixed  $\eta$-$\mu$ and $\kappa$-$\mu$ fading channels.
Closed-form expressions are obtained for the average symbol error rate (ASER) for general order rectangular QAM and cross QAM schemes using moment generating function based approach. Derived expressions are in the form of  Lauricella's $(F_D^{(n)}(\cdot), \Phi_1^{(n)}(\cdot))$ hypergeometric functions  which can be numerically evaluated using either integral or series representation. The obtained ASER expressions include other mixed fading channel cases addressed in the literature as special cases such as mixed Hoyt, and Rice fading,  mixed Nakagami-$m$, and Rice fading.  We further obtain a simple expression for the asymptotic ASER, which is useful to determine a factor governing the system performance at high SNRs, i.e., the diversity order.  Additionally, we analyze the optimal power allocation, which provides a practical design rule to optimally distribute the total transmission power between the source and the relay to minimize the ASER. Extensive numerical and computer simulation results are presented that confirm the accuracy of presented mathematical analysis.
\end{abstract}
\begin{IEEEkeywords}
Quadrature amplitude modulation, average symbol error rate, decode-and-forward, dual-hop relaying, mixed fading, optimal power allocation.
\end{IEEEkeywords}
\section{Introduction}
Cooperative communication has received considerable attention for several emerging wireless network architectures, such as cellular networks, local area
networks and heterogeneous networks to counter the effect of multipath fading which degrades the symbol error rate (SER) performance \cite{Sendonaris}-\cite{Abouelseoud}. The
cooperative communication offers significant increase in capacity and multiplexing gain
in the aforementioned wireless networks where user systems and nodes in a wireless
network share their resources and create collaboration through
distributed transmission. The motivation behind the cooperative
communication is to provide virtual multiple-input multiple-output
(MIMO) support for a user system which cannot accommodate multiple antenna due to size, complexity, power consumption and many other constraints. In a cooperative communication system, user terminals work as information sources as well as relays. The two main relaying methods used in cooperative communication scheme are amplify-and-forward (AF) and decode-and-forward (DF). In the AF case, the relay amplifies a signal transmitted by
the source and then re-transmits the same to the destination. In the DF case, the relay decodes a signal transmitted by the source and then encodes and transmits to the destination.

In recent years, there has been an increased interest on the performance analysis of dual-hop cooperative communication networks where links are subject to asymmetric fading conditions \cite{Duong}-\cite{Pawan} in which the propagation
in one of the links is dominated by a strong line-of-sight
(LOS) component relative to other links in the channel \cite{Kyosti}. The performance
analysis of dual-hop AF relay networks over mixed fading conditions are available in \cite{Duong}-\cite{PeppasAF} and the analysis
for  dual-hop DF relay networks over mixed fading conditions are studied in \cite{Kapucu}-\cite{Pawan}. In \cite{Pawan}, authors present exact SER expressions of $M$-ary phase shift keying (MPSK) scheme for DF relay system over $\kappa$-$\mu$ and $\eta$-$\mu$ and mixed $\kappa$-$\mu$ and $\eta$-$\mu$ fading channels. However, no closed-form SER expression of  MPSK in mixed $\kappa$-$\mu$ and $\eta$-$\mu$ fading channels has been provided.
There are a number of works reported on the performance of the DF cooperative diversity over fading channels focusing  on the capacity, diversity gain, and outage behavior of the system.
Past few years has seen a research interest on the investigation of error performance of various modulation schemes for the DF cooperative diversity over general fading channels. In \cite{Su} and \cite{Sadek}, SER performance for DF cooperation is examined over Rayleigh fading channels. The SER performance of different modulation schemes for dual-hop DF cooperation systems have been derived over Nakagami-$m$ fading channels in \cite{Ikki}-\cite{ShiDF}.
In \cite{Ikki}, the end-to-end bit error rate (BER)
performance of cooperative diversity networks using DF relaying over independent non-identical flat Nakagami-$m$ fading channels is available.
The SER expressions of MPSK and quadrature amplitude modulation (QAM) schemes for DF cooperative communications over Nakagami-$m$ are derived in \cite{Lee}.
In \cite{KimDF}, average BER performance of DF cooperative systems is investigated for binary PSK (BPSK) signals in Nakagami-$m$ fading channels for integer values of $m$.
Error probability of opportunistic DF relaying in Nakagami-$m$ fading channels with arbitrary $m$ is given in \cite{ShiDF}.
The authors in \cite{Yang} present the outage and error rate performance analysis of the DF cooperative diversity system with OSTBC over spatially correlated Nakagami-$m$ fading channels for integer values of $m$. The error performance of  MPSK and square QAM (SQAM) signals for dual-hop DF cooperation systems have been investigated over two-wave with diffuse power fading channels in \cite{Lu}.
In \cite{Fikadu}, authors analyze the performance of SQAM schemes with DF relaying system over Hoyt (Nakagami-$q$) fading channels.  In \cite{Muller}, the exact SER of $M$-ary PSK for multihop communication systems with regenerative relays is provided. In \cite{CaoMDF}, the outage probability (OP), SER, level crossing rate and average outage duration are derived for multihop DF cooperation systems over Generalized-$K$ fading channels.
The OP and BER of Gray-coded rectangular QAM (RQAM) signals for multihop DF cooperative systems have been analyzed  over $\eta$-$\mu$ fading channels in \cite{DixitDF}. A closed-form expression for the OP of dual-hop DF relaying in dissimilar Rayleigh fading channels is derived in \cite{BeaulieuDF}.
The authors in \cite{SuraweeraDF1} and \cite{Datsikas} are investigated the OP for the dual-hop DF protocol over Nakagami-$m$ fading channels.
The OP and outage capacity of dual-hop DF relaying system over $\eta$-$\mu$ and
$\kappa$-$\mu$ channels are given in \cite{Li1} and \cite{Li2}, respectively.

It is well known  that
the classical small-scale fading models (e.g., Rice, Hoyt, Nakagami-$m$) are not able to accurately characterize the small-scale variation of the fading signals, particularly at the tail probability \cite{Yacoub}.  Yacoub addressed this problem  suggesting two new distributions for fading models namely $\eta$-$\mu$ and $\kappa$-$\mu$ which better accommodates practical situations for which classical distributions are a poor fit. The $\eta$-$\mu$ distribution accurately models small-scale fading for various non-LOS (NLOS) conditions including the Hoyt, and Nakagami-$m$ distributions as special cases. On the other hand, the $\kappa$-$\mu$ distribution is a generalized distribution for modeling a great variety of LOS
channels and includes Rice and Nakagami-$m$ distributions as special cases.
Some available works on  performance analysis issues of cooperative systems operating in these two generic fading channel models are available in \cite{PeppasAF, Pawan, DixitDF, Li1, Li2} where the performance
of AF and DF cooperative systems are investigated.

QAM is a promising modulation scheme for a digital multimedia transmission in wireless communications due to its high bandwidth efficiency \cite{Proakis}. RQAM, SQAM and cross QAM (XQAM) are well known QAM schemes among which RQAM is a generic modulation scheme which includes SQAM, binary PSK (BPSK), orthogonal binary frequency-shift keying, quadrature PSK (QPSK) and multilevel amplitude shift-keying modulation schemes as  special cases. XQAM is an optimal QAM constellation for odd number of bits per symbol as it has low average symbol energy than RQAM \cite{Smith}. XQAM finds application  with constellations from $5$ bits to $15$ bits with usage in asymmetric digital subscriber lines and very high speed digital subscriber lines. Specifically, $32$ and $128$-XQAMs are employed in digital video broadcasting-cable. Moreover, XQAMs have been found to be useful in blind equalization \cite{Abrar} and adaptive modulation \cite{Panigrahi}, \cite{Zwingelstein}.

Despite many research works dealing with the DF cooperative diversity during the
past few years, to the best of our knowledge, closed-form expressions for the average SER (ASER) of general order RQAM and XQAM schemes with dual-hop DF relaying systems over mixed $\eta$-$\mu$ and $\kappa$-$\mu$ fading channels is not available in literature.
In this paper, we derive novel closed-form expressions for the ASER for general order RQAM and XQAM schemes
 with dual-hop DF relaying systems operating over mixed $\eta$-$\mu$ and $\kappa$-$\mu$ fading channels.
The obtained ASER expressions contain special functions such as Lauricella's  $(F_D^{(n)}(\cdot), \Phi_1^{(n)}(\cdot))$ hypergeometric functions which can be easily evaluated numerically using their finite integral or converging infinite
series representation \cite{Yu}-\cite{Martinez}. It is worth mentioning here that the derived expressions include the ASER expressions for mixed Hoyt, and Rice fading,  mixed Nakagami-$m$ and Rice fading, and non-identical Nakagami-$m$ fading as special cases. We further derive a simple expression for the asymptotic ASER, which is useful to determine a factor governing the system performance at high SNRs, i.e., the diversity order.  Additionally, we obtain the optimal power allocation factor, which provides a practical design rule to optimally distribute the total transmission power between the source and the relay to minimize the ASER.

The rest of the paper is organized as follows. Section II deals with system and channel models
and the ASER expressions are derived in Section III.
In Section IV, numerical results and discussion are given. The paper is concluded in Section V.

\section{System and Channel Models}
We consider a three-node system model \cite{Lee}.  
This model finds a typical application of cooperative communications for the uplink of cellular wireless systems.
This type of cooperative communication is also suitable for handsets equipped with single antenna for achieving transmit spatial diversity for link-quality improvement.
In this system model, the source node $s$ sends its information to  the destination node $d$ on two consecutive time slots.
In the first time slot, node $s$ broadcasts its symbol to node $d$ and relay node $r$.
In the second time slot, only node $r$, if decodes successfully, forwards the received symbol to node $d$. In this paper, we assume that the relay can successfully decode if the SNR level of the $sr$ link is above a predetermined threshold.  In a single relay dual-hop DF relaying system, node $i$ sends its information to node $j$  over $ij$ link, where $i\in \{s, r\}$, $j\in \{r, d\}$ and $ij\in\{sd, sr, rd\}$. When unit energy symbol $x$ is transmitted from node $i$, the baseband signal received at node $j$ can be given as

\begin{align}\label{yij}
  y_{ij}&=\sqrt{\tilde{P}_i}\,\alpha_{ij}\,x+n_{ij}
\end{align}
where $\tilde{P}_i$ is the power of the transmitted signal at node $i$, $\alpha_{ij}$
denotes fading channel coefficient of $ij$ link which is modeled as either $\eta$-$\mu$ or $\kappa$-$\mu$ distribution,  $n_{ij}$ is  additive white Gaussian noise (AWGN) with $N_0$ variance at node $j$. At node $s$ the symbol is transmitted with power $\tilde{P}_i=P_{s}$. If node $r$ successfully decodes the received symbol, the symbol is transmitted with power $\tilde{P}_i=P_{r}$, otherwise node $r$ remains idle, i.e. $P_r=0$. The total power $P$ of this dual-hop DF relaying system is given by $P=P_s+P_r$, $0<P_s, P_r<P$. Finally, node $d$ combines the signals of both the time slots
according to maximal ratio combining (MRC) technique.


The instantaneous SNR, $\gamma_{ij}$ of $ij$ link is defined as $\gamma_{ij}=\alpha_{ij}^2\tilde{P}_i/N_0$.
If a link is subjected to $\eta$-$\mu$ fading, the probability density function (PDF) of $\gamma_{ij}$ can be expressed as \cite{Yacoub}
\begin{align}\label{etamusnrpdf}
f_{\gamma_{ij}}(\gamma)&=\frac{2 \sqrt{\pi}\mu_{ij}^{\mu_{ij}+\frac{1}{2}} h_{ij}^{\mu_{ij}}\gamma^{\mu_{ij}-\frac{1}{2}}}{\Gamma(\mu_{ij})H_{ij}^{\mu_{ij}-\frac{1}{2}}\bar{\gamma}_{ij}^{\mu_{ij}+\frac{1}{2}}}
\exp\bigg(-\frac{2\mu_{ij} h_{ij} \gamma}{\bar{\gamma}_{ij}}\bigg)
I_{\mu_{ij}-\frac{1}{2}}\bigg(\frac{2\mu_{ij} H_{ij} \gamma}{\bar{\gamma}_{ij}}\bigg),
\end{align}
where $\Gamma(\cdot)$ is the gamma function, $I_{\nu}(\cdot)$ is the modified Bessel function of the first kind and $\nu$th order, $\mu_{ij}$ denotes the number of multipath clusters, and both $h_{ij}$ and $H_{ij}$ are functions of $\eta_{ij}$. In (\ref{etamusnrpdf}), $\bar{\gamma}_{ij}=\mathbb{E}[\gamma_{ij}]
=\Omega_{ij}\tilde{P}_i/N_0$ denotes the average SNR of the $ij$ link
, where $\mathbb{E}[\cdot]$ is the expectation operator and $\Omega_{ij}=\mathbb{E}[\alpha_{ij}^2]$ is the variance of $\alpha_{ij}$.
The distribution in (\ref{etamusnrpdf}) has been described for two types of physical models (Formats) depending on the way the parameter $\eta_{ij}$ is defined. In this paper, we consider only Format 1. In Format 1, $0<\eta_{ij}<\infty$ is the ratio of the power of inphase and quadrature phase components
of the scatter wave signals in each multipath cluster, with $h_{ij}=(2+\eta_{ij}^{-1}+\eta_{ij})/4$ and $H_{ij}=(\eta_{ij}^{-1}-\eta_{ij})/4$. It includes Hoyt ($\eta_{ij}=q^2$, $\mu_{ij}=0.5$ ), and Nakagami-$m$ ($\eta_{ij}=1$, $\mu_{ij}=m/2$) as special cases. The analysis presented here can also be extended to Format 2.

 If a link is subjected to $\kappa$-$\mu$ fading, the PDF of $\gamma_{ij}$ can be  expressed as \cite{Yacoub}
 \begin{align}\label{kappamusnrpdf}
f_{\gamma_{ij}}(\gamma)&=\frac{\mu_{ij}(1+\kappa_{ij})^{\frac{\mu_{ij}+1}{2}}\gamma^{\frac{\mu_{ij}-1}{2}}}
{\bar{\gamma}_{ij}^{\frac{\mu_{ij}+1}{2}}\kappa_{ij}^{\frac{\mu_{ij}-1}{2}}\exp(\mu_{ij}\kappa_{ij})}
 \exp\left(-\frac{(1+\kappa_{ij})\gamma}{\mu_{ij}^{-1}\bar{\gamma}_{ij}}\right)
\end{align}
where $\kappa_{ij}>0$ denotes the ratio of the total power due to  dominant components to
the total power due to scattered waves.
This fading  includes Rice ($\mu_{{ij}}=1$ and $\kappa_{ij}=K$), and
Nakagami-{$m$} ($\kappa_{ij}\rightarrow0$ and $\mu_{{ij}}=m$) as special cases.

Let $\gamma_{th}$ be the predetermined threshold for the $sr$ link .
If instantaneous SNR $\gamma_{sr}$ of $sr$ link falls below $\gamma_{th}$,
the link from node $s$ to node $d$ via node $r$ is assumed to fail. On the other hand, if node $r$ can successfully decode the messages from node $s$ and forwards to node $d$, then the PDF of  the end to end SNR $\gamma_{srd}$  can be given as \cite{CaoMDF}, \cite{BeaulieuDF}
  \begin{eqnarray}\label{e2epdf}
  f_{\gamma_{srd}}(\gamma)=A_{sr}\delta(\gamma)+(1-A_{sr})f_{\gamma_{rd}}(\gamma),
  \end{eqnarray}
  where $A_{sr}$ is the probability that outage occurs in $sr$ link and $\delta(\cdot)$ is Dirac delta function.
The probability $A_{sr}$ in (\ref{e2epdf}) can be calculated as \cite{CaoMDF}
  \begin{align}\label{A}
  A_{sr}&=1-\mbox{Pr}\{\gamma_{sr}>\gamma_{th}\}
  =F_{\gamma_{sr}}(\gamma_{th}),
  \end{align}
  where $\mbox{Pr}\{\cdot\}$ is the probability operation and $F_{\gamma_{sr}}(\cdot)$ denotes the cumulative distribution of function of $\gamma_{sr}$.

\section{Average Symbol Error Rate Analysis}
Mathematically, ASER $P(e)$  of any modulation scheme for considered system can be
 computed as \cite{Simonbook}
\begin{equation}\label{ASER}
P(e)= \int_{0}^{\infty} P(e|\gamma)f_{\gamma_t}(\gamma)d\gamma,
\end{equation}
where $P(e|\gamma)$ is the conditional SER of the modulation scheme in AWGN channels and
$f_{\gamma_t}(\gamma)$ is the the PDF of the total SNR $\gamma_t$ of output of the MRC combiner at node $d$.
To obtain the closed-form ASER expression, the moment generating function (MGF) of $\gamma_t$ is required.
The $\gamma_t$ can be given as \cite{Simonbook}
\begin{align}
\gamma_t=\gamma_{sd}+\gamma_{srd}.
\end{align}
The MGF of $\gamma_u$ is defined as \cite{Simonbook}
\begin{eqnarray}\label{mgfu}
\mathcal{M}_{\gamma_{u}}(z)
=\int_{0}^{\infty}\exp(-z\gamma)f_{\gamma_{u}}(\gamma)d\gamma,
\end{eqnarray}
where $u\in \{sd, rd, srd, t\}$, and $z$ is the Laplace variable.
The $\mathcal{M}_{\gamma_t}(z)$ of $\gamma_t$ can be obtained from the formula \cite{Simonbook}
\begin{align}\label{mgft1}
\mathcal{M}_{\gamma_t}(z)&=\mathcal{M}_{\gamma_{sd}}(z)\,\mathcal{M}_{\gamma_{srd}}(z).
\end{align}

Using (\ref{e2epdf}) in (\ref{mgfu}), the expression of $\mathcal{M}_{\gamma_{srd}}(z)$ can be obtained as \cite{ErmolovaRQAM}
\begin{eqnarray}\label{mgfsrd}
\mathcal{M}_{\gamma_{srd}}(z)=A_{sr}+(1-A_{sr})M_{\gamma_{rd}}(z),
\end{eqnarray}

Finally, substituting (\ref{mgfsrd}) in (\ref{mgft1}), $\mathcal{M}_{\gamma_t}(z)$ can be written as
\begin{align}
\mathcal{M}_{\gamma_t}(z)&=A_{sr}\,\mathcal{M}_{\gamma_{sd}}(z)+(1-A_{sr})M_{\gamma_{sd}}(z)\,\mathcal{M}_{\gamma_{rd}}(z).
\label{mgft2}
\end{align}
Using (\ref{etamusnrpdf}) in (\ref{mgfu}), the $\mathcal{M}_{\gamma_{ij}}(z)$ of $\eta$-$\mu$ faded $ij$ link can be obtained as \cite{ErmolovaRQAM}
\begin{eqnarray}\label{mgfsd}
\mathcal{M}_{\gamma_{ij}}(z)
 =\frac{(4\mu_{ij}^2h_{ij})^{\mu_{ij}}(2\mu_{ij}(h_{ij}+H_{ij})+z\bar{\gamma}_{ij})^{-\mu_{ij}}}
{(2\mu_{ij}(h_{ij}-H_{ij})+z\bar{\gamma}_{ij})^{\mu_{ij}}}.
\end{eqnarray}
Using (\ref{kappamusnrpdf}) in (\ref{mgfu}), the $\mathcal{M}_{\gamma_{ij}}(z)$ of $\kappa$-$\mu$ faded $ij$ link can be obtained as \cite{ErmolovaRQAM}
\begin{align}\label{mgfrd}
\mathcal{M}_{\gamma_{ij}}(z)
&=\left(\frac{\mu_{ij}(1+\kappa_{ij})}{\mu_{ij}(1+\kappa_{ij})+z\bar{\gamma}_{ij}}\right)^{\mu_{ij}}
 \exp\left(\frac{-z\mu_{ij}\kappa_{ij}\bar{\gamma}_{ij}}{\mu_{ij}(1+\kappa_{ij})+z\bar{\gamma}_{ij}}\right).
\end{align}

The scenario, where one hop of the link is subjected to $\eta$-$\mu$ fading and the other
hop is subjected to $\kappa$-$\mu$ fading, can be modeled accurately by mixed NLOS
and LOS conditions, which occur in various applications including micro-/macrocellular and/or hybrid satellite/terrestrial communication systems \cite{Soliman},\cite{Kyosti}. We assume that all links experience independent and non identical but asymmetric fading. With two different fading types, there are six possible combinations for mixed fading scenarios as given in Table \ref{Scenario}.
\begin{table}[h]
\begin{center}
\caption{Fading parameters of different scenarios.}
\label{Scenario}
\begin{tabular}{|c|c|c|c|}
\hline
\multirow{2}{*}{Scenario} & \multicolumn{3}{l|}{Links parameters}                                 \\ \cline{2-4}
& $sd$ & $sr$ & $rd$                  \\ \hline
1 & $\eta_{sd}, \mu_{sd}$ & $\eta_{sr}, \mu_{sr}$ & $\kappa_{rd}, \mu_{rd}$ \\ \hline
2 & $\kappa_{sd}, \mu_{sd}$ & $\eta_{sr}, \mu_{sr}$ & $\eta_{rd}, \mu_{rd}$ \\ \hline
3 & $\kappa_{sd}, \mu_{sd}$ & $\kappa_{sr}, \mu_{sr}$ & $\eta_{rd}, \mu_{rd}$ \\ \hline
4  & $\eta_{sd}, \mu_{sd}$ & $\kappa_{sr}, \mu_{sr}$ & $\kappa_{rd}, \mu_{rd}$ \\ \hline
5 & $\kappa_{sd}, \mu_{sd}$ & $\eta_{sr}, \mu_{sr}$ & $\kappa_{rd}, \mu_{rd}$ \\ \hline
6 & $\eta_{sd}, \mu_{sd}$ & $\kappa_{sr}, \mu_{sr}$ & $\eta_{rd}, \mu_{rd}$ \\ \hline
\end{tabular}
\end{center}
\end{table}
We have analyzed the mixed fading scenario 1 where $\gamma_{sd}$ and $\gamma_{sr}$ are modeled as $\eta$-$\mu$ distribution but $\gamma_{rd}$ is modeled as $\kappa$-$\mu$ distribution.
For $\eta$-$\mu$ faded $sr$ link
$A_{sr}$  can be given as  \cite{JimenezMRC}
  \begin{align}\label{etamucdf}
A_{sr}&=1-Y_{\mu_{sr}}\Bigg(\frac{H_{sr}}{h_{sr}},\sqrt{\frac{2\mu_{sr} h_{sr}\gamma_{th}}{\bar{\gamma}_{sr}}}\Bigg),
\end{align}
where $Y_\nu(a,b)=\frac{2^{\frac{3}{2}-\nu}\sqrt{\pi}(1-a^2)^{\nu}}{a^{\nu-\frac{1}{2}}\Gamma(\nu)}
\int_b^{\infty}x^{2\nu}\mbox{e}^{-x^2}I_{\nu-\frac{1}{2}}(a\, x^2)dx$ is the Yacoub's integral \cite{Yacoub}.
A general solution of this integral is  \cite{JimenezMRC}
\begin{align}\label{Yacoubintegral1}
&Y_\nu(a,b)=1-\frac{\Phi_2^{(2)}(\nu,\nu;1+2\nu;-(1+a)b^2,-(1-a)b^2)}{(1-a^2)^{-\nu} b^{-4\nu}\Gamma(1+2\nu)},
\end{align}
where $\Phi_2^{(2)}$ is the confluent Lauricella function \cite{Prudnikov4}. In \cite{JimenezMRC}, another closed-form expression for the integer values of $2\mu_{sr}$ is also presented.
\subsection{Exact Analysis}
\subsubsection{$M$-ary RQAM }
The conditional SER performance of $M=M_I\times M_Q$-ary RQAM in AWGN channels is given as  \cite{DixittwdpQAM}, \cite{BeaulieuRQAM}
\begin{align}\label{CSERRQAM2}
P(e|\gamma)&=2p\,Q\left(a\sqrt{\gamma},\pi/2\right)+2q\,Q\left(b\sqrt{\gamma},\pi/2\right)
-2p\,q\,\big\{Q\big(a\sqrt{\gamma},\text{arccot}(b/a)\big)
\nonumber \\ &
+Q\big(b\sqrt{\gamma},\arctan(b/a)\big\},
\end{align}
where $p=1-\frac{1}{M_I}$, $q=1-\frac{1}{M_Q}$, $M_I$ and $M_Q$ are the number of in-phase and quadrature-phase
constellation points, respectively,
$a=\sqrt{\frac{6}{(M_I^2-1)+(M_Q^2-1)\beta^2}}$, $b=\beta a$ and $\beta=d_Q/d_I$ is the quadrature-to-in-phase
decision distance ratio with $d_I$ and $d_Q$ being
the in-phase and quadrature decision distance, respectively, and $Q(x,\phi)$ is given as\cite{Yu}, \cite{Simonbook}
\begin{eqnarray}\label{qfunction1}
 Q(x,\phi)=\frac{1}{\pi}\int_{0}^{\phi}\exp\Big(-\frac{x^2}{2\sin^2\theta}\Big)d\theta; \qquad x\geq 0.
\end{eqnarray}
Substituting (\ref{CSERRQAM2}) in (\ref{ASER}) and by algebraic manipulations  ASER for
RQAM scheme denoted as  $P^{\text{RQAM}}(e)$, can be expressed as
\begin{align}\label{ASERRQAM}
&P^{\text{RQAM}}(e)
=2p\,\mathcal{I}\big(a,\pi/2\big)+2q\,\mathcal{I}\big(b,\pi/2\big)
-2p\,q\big\{\mathcal{I}\big(a,\text{arccot}(b/a)\big)
+\mathcal{I}\big(b,\arctan(b/a)\big)\big\},
\end{align}
where the function $\mathcal{I}\big(\cdot,\cdot)$ is defined as
\begin{align}\label{I}
\mathcal{I}\big(x,\phi\big)&=\int_{0}^{\infty}Q(x\sqrt{\gamma},\phi)f_{\gamma_t}(\gamma)d\gamma
\nonumber \\ & =\frac{1}{\pi}\int_0^\phi \int_{0}^{\infty}\exp\Big(-\frac{x^2\gamma}{2\sin^2\theta}\Big) f_{\gamma_t}(\gamma)d\gamma d\theta
\nonumber\\ &=\frac{1}{\pi}\int_0^\phi\mathcal{M}_{\gamma_t}\Big(\frac{x^2}{2\sin^2\theta}\Big)d\theta.
\end{align}
Now, substituting (\ref{mgft2}) in (\ref{I}), $\mathcal{I}(x,\phi)$ can be written as
\begin{align}\label{I12}
\mathcal{I}\big(x,\phi\big)
&=A_{sr}\,\mathcal{I}_1\big(x,\phi\big)+(1-A_{sr})\,\mathcal{I}_2\big(x,\phi\big),
\end{align}
where
\begin{align}\label{I1}
\mathcal{I}_1\big(x,\phi\big)&=\frac{1}{\pi}\int_0^\phi\mathcal{M}_{\gamma_{sd}}
\Big(\frac{x^2}{2\sin^2\theta}\Big)d\theta,
\end{align}
and
\begin{align}\label{I2}
\mathcal{I}_2\big(x,\phi\big)&=\frac{1}{\pi}
\int_0^\phi\mathcal{M}_{\gamma_{sd}}\Big(\frac{x^2}{2\sin^2\theta}\Big)
\mathcal{M}_{\gamma_{rd}}\Big(\frac{x^2}{2\sin^2\theta}\Big)d\theta.
\end{align}
The closed-form expressions for $\mathcal{I}_1(x, \phi)$ and $\mathcal{I}_2(x, \phi)$ are derived in Appendix. Thus, the ASER expression for $P^{\text{RQAM}}(e)$ can be given as
\begin{align}\label{ASERRQAM_Final}
&P^{\text{RQAM}}(e)
=A_{sr}\big\{2p\,\mathcal{I}_1\big(a,\pi/2\big)+2q\,\mathcal{I}_1\big(b,\pi/2\big)
-2p\,q\mathcal{I}_1\big(a,\text{arccot}(b/a)\big)
-2p\,q\mathcal{I}_1\big(b,\arctan(b/a)\big)\big\}
\nonumber\\&+(1-A_{sr})\big\{2p\,\mathcal{I}_2\big(a,\pi/2\big)+2q\,\mathcal{I}_2\big(b,\pi/2\big)
-2p\,q\mathcal{I}_2\big(a,\text{arccot}(b/a)\big)
-2p\,q\mathcal{I}_2\big(b,\arctan(b/a)\big)\big\}.
\end{align}

For the special case of  $M$-ary SQAM,\footnote{Substitute $y=x$ into (\ref{Ix12}) and (\ref{Ix22}) to obtain $\mathcal{I}_1\left(x,\pi/4\right)$ and $\mathcal{I}_2\left(x,\pi/4\right)$, respectively} i.e.  when $M_I = M_Q = \sqrt{M}$ and $\beta = 1$, it can be shown that
(\ref{ASERRQAM_Final}) reduces to
\begin{align}\label{ASERSQAM_Final}
&P^{\text{SQAM}}(e)
=4A_{sr}\,\widetilde{p}\big\{\mathcal{I}_1\big(\widetilde{a},\pi/2\big)
-\widetilde{p}\,\mathcal{I}_1\big(\widetilde{a},\pi/4\big)
\big\}
+4(1-A_{sr})\,\widetilde{p}\big\{\mathcal{I}_2\big(\widetilde{a},\pi/2\big)
-\widetilde{p}\,\mathcal{I}_2\big(\widetilde{a},\pi/4\big)
\big\},
\end{align}
where $\widetilde{p}=1-\frac{1}{\sqrt{M}}$ and $\widetilde{a}=\sqrt{\frac{3}{M-1}}$.
For the special case of  BPSK, i.e.  when $M_I = 2$, $M_Q = 1$ and $\beta = 0$, it can be shown that
(\ref{ASERRQAM_Final}) reduces to

\begin{align}\label{ASERBPSK_Final}
P^{\text{BPSK}}(e)
&=A_{sr}\,\mathcal{I}_1\big(\sqrt{2},\pi/2\big)+(1-A_{sr})\,\mathcal{I}_2\big(\sqrt{2},\pi/2\big).
\end{align}
\subsubsection{$M$-ary XQAM }
The conditional SER performance of $M$-ary XQAM in AWGN channels can be  given   as \cite{Yu}
\begin{align}\label{CSERXQAM}
P(e|\gamma)
&=g_1Q\big(a_0\sqrt{\gamma},\pi/2\big)
-g_2\sum_{k=1}^{\nu-1}Q(a_k\sqrt{\gamma},\arctan(k/(k+1)))
\nonumber\\&+g_2Q\big(a_1\sqrt{\gamma},\pi/2\big)+g_2\sum_{k=2}^{\nu}Q(a_k\sqrt{\gamma},\arctan(k/(k-1)))
\nonumber\\&-g_3Q(a_0\sqrt{\gamma},\pi/4)
-2g_2\sum_{k=1}^{\nu-1}Q(a_0\sqrt{\gamma},\arctan(1/(2k+1))),
\end{align}

where $M=32, 128, 512,\ldots$, $g_1=4-\frac{6}{\sqrt{2M}}$, $g_2=\frac{4}{M}$, $g_3=4-\frac{12}{\sqrt{2M}}+\frac{12}{M}$, $\nu=\frac{\sqrt{2M}}{8}$, $a_0=\sqrt{\frac{96}{(31M-32)}}$,
$a_k=\sqrt{2}k a_0, k=1,2, \ldots, \nu$.

Using a similar approach as followed to obtained (\ref{ASERRQAM_Final}), a
closed-form ASER expression of XQAM can be obtained as given
\begin{align}\label{ASERXQAM_Final1}
&P^{\text{XQAM}}(e)=(1-A_{sr})\Big\{\mathcal{I}_2\big(a_0,\pi/2\big)
+g_2\mathcal{I}_2\big(a_1,\pi/2\big)\Big\}
+A_{sr}\Big\{\mathcal{I}_1\big(a_0,\pi/2\big)
\nonumber\\&
-g_2\sum_{k=1}^{\nu-1}\mathcal{I}_1\big(a_k,\arctan(k/(k+1))\big)
+g_2\mathcal{I}_1\big(a_1,\pi/2\big)+g_2\sum_{k=2}^{\nu}\mathcal{I}_1\big(a_k,\arctan(k/(k-1))\big)
\nonumber\\&-g_3\mathcal{I}_1\big(a_0,\pi/4\big)
-2g_2\sum_{k=1}^{\nu-1}\mathcal{I}_1\big(a_0,\arctan(1/(2k+1))\big)\Big\}
+(1-A_{sr})\Big\{-g_3\mathcal{I}_2\big(a_0,\pi/4\big)
\nonumber\\&
g_2\sum_{k=2}^{\nu}\mathcal{I}_2\big(a_k,\arctan(k/(k-1))\big)
-2g_2\sum_{k=1}^{\nu-1}\mathcal{I}_2\big(a_0,\arctan(1/(2k+1))\big)
\nonumber\\&
-g_2\sum_{k=1}^{\nu-1}\mathcal{I}_2\big(a_k,\arctan(k/(k+1))\big)
\Big\}.
\end{align}
\subsection{Asymptotic Analysis}
In asymptotic analysis, we assume that the average SNR values for three links are sufficiently large i.e., $\bar{\gamma}_{ij}>>1$. With the high SNR assumption, the MGF of $\eta$-$\mu$ faded link in (\ref{mgfsd}) can be approximated as
\begin{equation}\label{mgfsdapprox}
\mathcal{M}_{\gamma_{ij}}^{\infty}(z)
 \simeq \left(\frac{2\mu_{ij}\sqrt{h_{ij}}}{z\,\bar{\gamma}_{ij}}\,\right)^{2\mu_{ij}},
 \end{equation}
and the MGF of $\kappa$-$\mu$ faded link in (\ref{mgfrd}) can be approximated as
\begin{equation}\label{mgfrdapprox}
\mathcal{M}_{\gamma_{ij}}^{\infty}(z)
 \simeq \left(\frac{\mu_{ij}(1+\kappa_{ij})}{z\,\bar{\gamma}_{ij}}\right)^{\mu_{ij}}
  \exp\left(-\mu_{ij}\kappa_{ij}\right).
\end{equation}
Thus, $A_{sr}$ of $sr$ link can be approximated as
\begin{eqnarray}\label{Aapprox1}
    A_{sr}^{\infty}&\simeq& \mathcal{L}^{-1}\left\{\frac{\mathcal{M}_{\gamma_{sr}}^{\infty}(z)}{z}\right\}\bigg|_{\gamma={\gamma_{th}}}
		\\ \nonumber
		&\simeq&\frac{1}{\Gamma(2\mu_{sr}+1)} \left(\frac{2\mu_{sr}\sqrt{h_{sr}}\gamma_{th}}{\bar{\gamma}_{sr}}\,\right)^{2\mu_{sr}},
\end{eqnarray}
where $\mathcal{L}^{-1}\{\cdot\}$ is the inverse Laplace transform operator.
For high SNR, we can also assume
 $1-A_{sr}^{\infty}\simeq 1$.
\subsubsection{$M$-ary RQAM}
Using (\ref{mgfsdapprox}), (\ref{mgfrdapprox}) and (\ref{Aapprox1}), the asymptotic ASER for RQAM,
 $P_{\infty}^{\text{RQAM}}(e)$ can be expressed as
\begin{align}\label{ASERRQAM_Finalapprox}
P^{\text{RQAM}}_{\infty}(e)
&=A_{sr}^{\infty}\big\{2p\,\mathcal{I}_1^{\infty}\big(a,\pi/2\big)+2q\,\mathcal{I}_1^{\infty}\big(b,\pi/2\big)
-2p\,q\mathcal{I}_1^{\infty}\big(a,\text{arccot}(b/a)\big)
\nonumber\\&
-2p\,q\mathcal{I}_1^{\infty}\big(b,\arctan(b/a)\big)\big\}
-2p\,q\mathcal{I}_2^{\infty}\big(a,\text{arccot}(b/a)\big)
-2p\,q\mathcal{I}_2^{\infty}\big(b,\arctan(b/a)\big)
\nonumber\\&
+2p\,\mathcal{I}_2^{\infty}\big(a,\pi/2\big)+2q\,\mathcal{I}_2^{\infty}\big(b,\pi/2\big),
\end{align}
 where
\begin{align}\label{I1sdapprox}
\mathcal{I}_1^{\infty}\left(x,\phi\right)&=\frac{1}{\pi}\int_0^\phi\mathcal{M}_{\gamma_{sd}}^{\infty}
\Big(\frac{x^2}{2\sin^2\theta}\Big)d\theta,
\end{align}
and
\begin{align}\label{I2srdapprox}
&\mathcal{I}_2^{\infty}\left(x,\phi\right)
=\frac{1}{\pi}
\int_0^\phi\mathcal{M}_{\gamma_{sd}}^{\infty}\Big(\frac{x^2}{2\sin^2\theta}\Big)
\mathcal{M}_{\gamma_{rd}}^{\infty}\Big(\frac{x^2}{2\sin^2\theta}\Big)d\theta.
\end{align}
The closed-form expressions for $\mathcal{I}_1^{\infty}\left(x,\phi\right)$ and $\mathcal{I}_2^{\infty}\left(x,\phi\right)$ are derived in Appendix.
\subsubsection{$M$-ary XQAM}
Using an approach similar to that followed to obtain (\ref{ASERRQAM_Finalapprox}), a
closed-form asymptotic ASER expression for XQAM can be given as
\begin{align}\label{ASERXQAM_Final1approx}
&P^{\text{XQAM}}_{\infty}(e)
=A_{sr}^{\infty}\Big\{\mathcal{I}_1^{\infty}\big(a_0,\pi/2\big)
-g_2\sum_{k=1}^{\nu-1}\mathcal{I}_1^{\infty}\big(a_k,\arctan(k/(k+1))\big)
\nonumber\\&+g_2\mathcal{I}_1^{\infty}\big(a_1,\pi/2\big)
+g_2\sum_{k=2}^{\nu}\mathcal{I}_1^{\infty}\big(a_k,\arctan(k/(k-1))\big)
\nonumber\\&-g_3\mathcal{I}_1^{\infty}\big(a_0,\pi/4\big)
-2g_2\sum_{k=1}^{\nu-1}\mathcal{I}_1^{\infty}\big(a_0,\arctan(1/(2k+1))\big)\Big\}
\nonumber\\&+\Big\{\mathcal{I}_2^{\infty}\big(a_0,\pi/2\big)
-g_2\sum_{k=1}^{\nu-1}\mathcal{I}_2^{\infty}\big(a_k,\arctan(k/(k+1))\big)
\nonumber\\&+g_2\mathcal{I}_2^{\infty}\big(a_1,\pi/2\big)
+g_2\sum_{k=2}^{\nu}\mathcal{I}_2^{\infty}\big(a_k,\arctan(k/(k-1))\big)
\nonumber\\&-g_3\mathcal{I}_2^{\infty}\big(a_0,\pi/4\big)
-2g_2\sum_{k=1}^{\nu-1}\mathcal{I}_2^{\infty}\big(a_0,\arctan(1/(2k+1))\big)\Big\}.
\end{align}
The asymptotic ASER expressions in (\ref{ASERRQAM_Finalapprox}) and (\ref{ASERXQAM_Final1approx}) can be rewritten by substituting $P_s=\xi P$ and $P_r=(1-\xi)P$ for $0\leq\xi\leq1$, as \begin{equation}\label{PeAsym}
    P_{\infty}^{\text{QAM}}(e)=\mathcal{E}_1\left(\frac{P}{N_0}\right)^{-2(\mu_{sd}+\mu_{sr})}
    +\mathcal{E}_2\left(\frac{P}{N_0}\right)^{-(2\mu_{sd}+\mu_{rd})}
\end{equation}
where $\mathcal{E}_1$ and $\mathcal{E}_2$ summarize the necessary elements in the asymptotic ASER expressions. The $\xi$ is optimal power allocation factor.
\subsection{Diversity Order Analysis}
The diversity order, $\mathcal{O}_{\text{DF}}$ of dual-hop DF relaying system can be calculated as \cite{Song}
\begin{align}\label{DO}
    \mathcal{O}_{\text{DF}}&=\lim_{\frac{P}{N_0}\rightarrow \infty}\frac{-\log P^{\text{QAM}}(e)}{\log \frac{P}{N_0} }
\end{align}
Using (\ref{PeAsym}) and (\ref{DO}), we can show that our considered scenario 1 (in Table~\ref{Scenario}) of this system achieves the maximum diversity order of $\mathcal{O}_{\text{DF}}=2\mu_{sd}+\min\{2\mu_{sr},\mu_{rd}\}$. It can be observed that the fading parameters $\eta_{sd}, \eta_{sr}$ and $\kappa_{rd}$ have no impact on the diversity order.
\begin{figure}[t]
\begin{center}
\includegraphics[width=12.4cm,height=8.6cm]{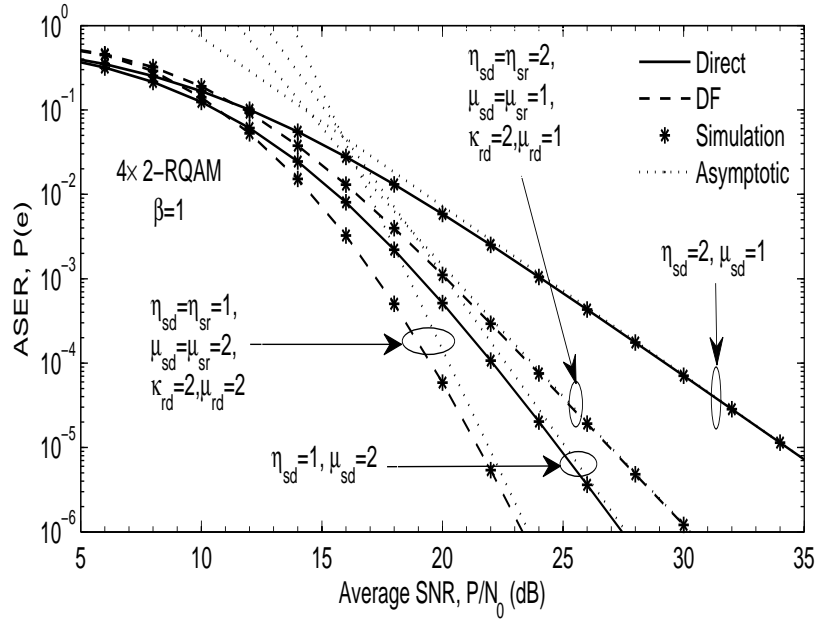}
\caption{ASER of $4\times 2$-RQAM scheme with varying $\eta$, $\kappa$ and $\mu$.}
\label{f2}
\end{center}
\end{figure}

\begin{figure}[t]
\begin{center}
\includegraphics[width=12.4cm,height=8.6cm]{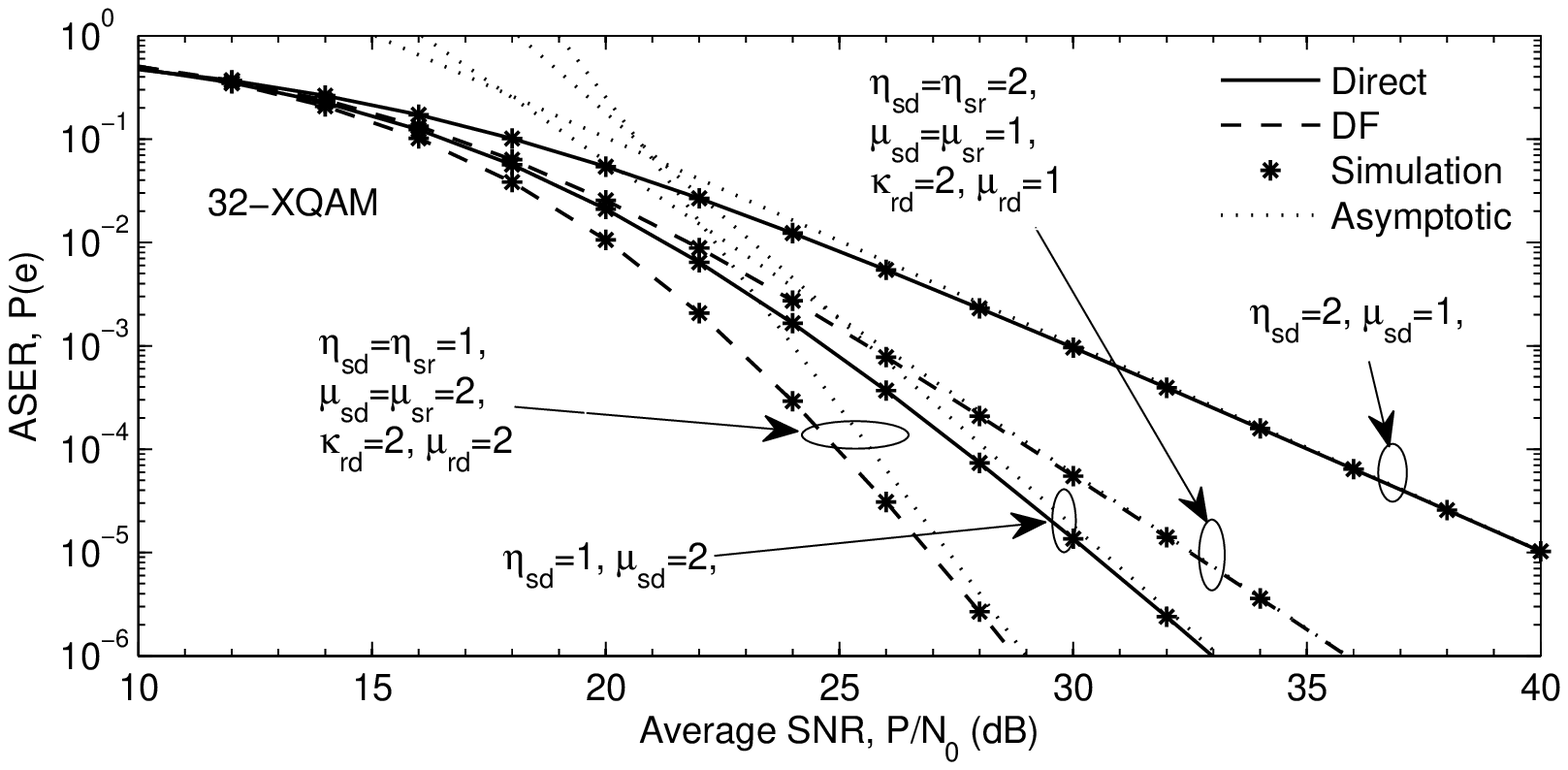}
\caption{ASER of $32$-XQAM scheme with varying $\eta$, $\kappa$ and $\mu$.}
\label{f3}
\end{center}
\end{figure}
\subsection{Optimal Power Allocation Analysis}
Equal power allocation strategy for transmitters is not an optimal solution for power allocation. If partial channel state information is available at the transmitting nodes, an optimal power allocation can be performed to minimize the ASER at the destination node\cite{Lee}. The asymptotic ASER expressions in (\ref{ASERRQAM_Finalapprox}) and (\ref{ASERXQAM_Final1approx}) can be rewritten by substituting $P_s=\xi P$ and $P_r=(1-\xi)P$ for $0\leq\xi\leq1$,  as

\begin{align}\label{Ps}
P_{\infty}^{\text{QAM}}(\xi)
&=\frac{\Omega_{sd}^{-2\mu_{sd}}\Omega_{sr}^{-2\mu_{sr}}\mathcal{C}_1}{\xi^{2(\mu_{sd}+\mu_{sr})}}+
\frac{\Omega_{sd}^{-2\mu_{sd}}\xi^{-2\mu_{sd}}\mathcal{C}_2}{\Omega_{rd}^{\mu_{rd}}(1-\xi)^{\mu_{rd}}},
    \end{align}
where $\mathcal{C}_1, \mathcal{C}_2\geq 0$ are constants of the expression\footnote{not given here as it is not required for optimization}. It can be shown that the second derivative of (\ref{Ps}) w.r.t $\xi$ is always grater than or equal to 0, i.e., $d^2 P_{\infty}^{\text{QAM}}(\xi)/d \xi^2\geq 0$. Hence (\ref{Ps}) is a convex function w.r.t. $\xi$. Equating the first derivative of (\ref{Ps}) w.r.t. $\xi$ to zero, we get the relation
\begin{align}\label{Pszero}
     -2\mathcal{C}_1(\mu_{sd}+\mu_{sr})\Omega_{rd}^{\mu_{rd}}(1-\xi)^{\mu_{rd}+1}
     &=\mathcal{C}_2\Omega_{sr}^{2\mu_{sr}}\xi^{2\mu_{sr}}(2\mu_{sd}-(2\mu_{sd}+\mu_{rd})\xi).
\end{align}
It can be observed from (\ref{Pszero}) that the asymptotic optimum power allocation does not depend on the variance of link between $s$ and $d$, it depends only on the variances of links between $s$ and $r$ and, between $r$ and $d$.  Moreover, we can also infer that the optimum transmitted power $P_s$ at $s$ is larger than
$2\mu_{sd}P/(2\mu_{sd}+\mu_{rd})$ and less than $P$, while the optimum power $P_r$ used at $r$ is larger than $0$ and less than $ \mu_{rd}P/(2\mu_{sd}+\mu_{rd})$. It means that we should always allocate more power at $s$ and less power at $r$.
\section{Numerical Results and Discussion}
Numerical examples concerning the ASER performance of QAM schemes are presented
with computer simulations to verify the accuracy of analytical results. The Lauricella's hypergeometric functions are numerically evaluated using their finite integral representation.
In the numerical evaluation, we assume that the power is equally allocated to the source and the relay and the variance of each link is unity, i.e., $\Omega_{sd}=\Omega_{sr}=\Omega_{rd}=1$, unless stated otherwise.  Additionally,
the threshold $\gamma_{th}$ is chosen according
to $\gamma_{th}=2^{2R}-1$, with the spectrum efficiency $R$
being set to $1$ bit/s/Hz.
\begin{figure}[t]
\begin{center}
\includegraphics[width=12.4cm,height=8.6cm]{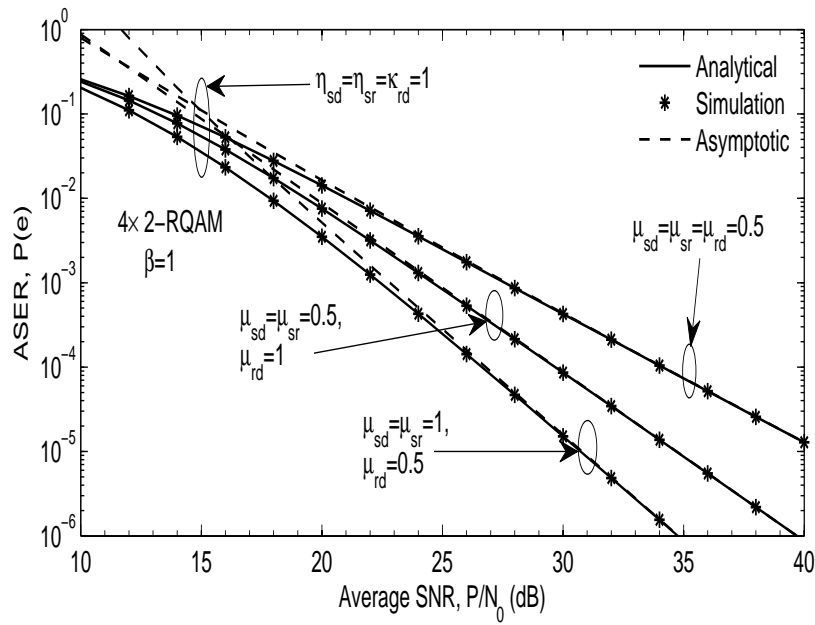}
\caption{ASER of $4\times 2$-RQAM scheme for dual-hop DF relaying system with varying $\mu$ and fixed $\eta=\kappa=1$.}
\label{f4}
\end{center}
\end{figure}
\begin{figure}[t]
\begin{center}
\includegraphics[width=12.4cm,height=8.6cm]{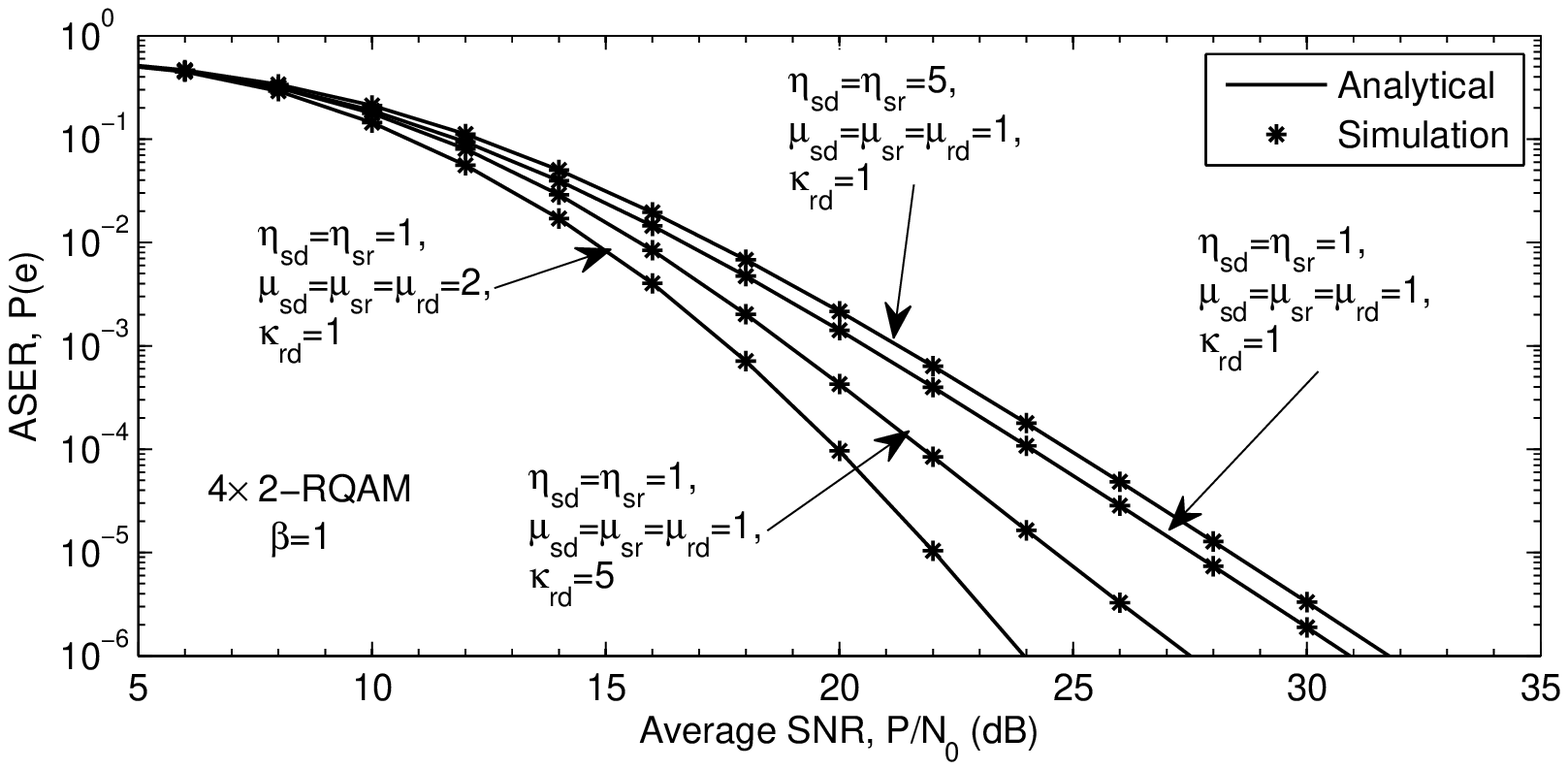}
\caption{ASER of $4\times 2$-RQAM scheme for dual-hop DF relaying system with varying $\eta$, $\kappa$ and $\mu$.}
\label{f5}
\end{center}
\end{figure}
\begin{figure}[t]
\begin{center}
\includegraphics[width=12.4cm,height=8.6cm]{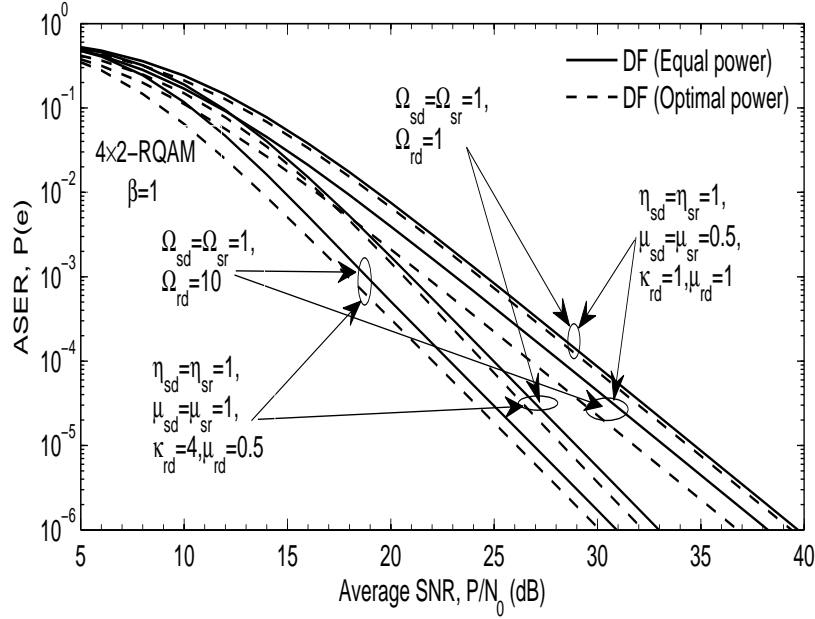}
\caption{ASER of $4\times2$-RQAM scheme for dual-hop DF relaying system with equal and optimal allocation.}
\label{f6}
\end{center}
\end{figure}
\subsection{Direct Communication versus DF Cooperative Communication with RQAM and XQAM Schemes }
Figs. \ref{f2}, and \ref{f3} illustrate the ASER for $4\times 2$-RQAM, and $32$-XQAM schemes, respectively.
We notice that the numerical results match with the corresponding computer simulation results.
These figures also show that the DF cooperative communication can always substantially improve the ASER performance, relative to the direct communication over the medium-to-high SNR region. For example, it can be
observed from Fig.~\ref{f2} that for an ASER of $10^{-4}$, $P/N_0$
improvement that can be achieved in dual-hop DF relaying system ($\eta_{sd}=\eta_{sr}=1,\,\mu_{sd}=\mu_{sr}=2,\,\kappa_{rd}=2,\,\mu_{rd}=2$) over
direct communication ($\eta_{sd}=1,\,\mu_{sd}=2$) is $3$dB (approx.)
\subsection{Impact of Fading Parameters}
Fig.~\ref{f4} shows the impact of $\mu_{sd}=\mu_{sr}$, and $\mu_{rd}$ for fixed $\eta_{sd}=\eta_{sr}=\kappa_{rd}=1$ on the ASER of $4\times 2$-RQAM scheme for dual-hop DF relaying system over mixed $\eta$-$\mu$ and $\kappa$-$\mu$ fading channels. It can be observed from this figure that the SNR gains are achieved resulting from increasing the fading parameters, $\mu_{sd}=\mu_{sr}$, and $\mu_{rd}$. For example, an ASER of $10^{-4}$ with $\eta_{sd}=\eta_{sr}=\kappa_{rd}=1$ occurs at $P/N_0\approx 34$dB  when $\mu_{sd}=\mu_{sr}=\mu_{rd}=0.5$, $P/N_0\approx 30$dB  when $\mu_{sd}=\mu_{sr}=0.5$, and $\mu_{rd}=1$, and, $P/N_0\approx 26.5$dB  when $\mu_{sd}=\mu_{sr}=1$, and $\mu_{rd}=0.5$.

Fig.~\ref{f5} shows the impact of $\eta$, $\kappa$ and $\mu$ on the ASER of $4\times 2$-RQAM scheme for dual-hop DF relaying system.
Two important observations can be drawn from this figure. First, the SNR gains are resulting from increasing the fading parameters, $\kappa$ and $\mu$, as expected. For example, an ASER of $10^{-4}$ occurs at $P/N_0\approx 24$dB  when $\eta_{sd}=\eta_{sr}=1$, $\mu_{sd}=\mu_{sr}=\mu_{rd}=1$, and $\kappa_{rd}=1$, $P/N_0\approx 22$dB  when $\eta_{sd}=\eta_{sr}=1$, $\mu_{sd}=\mu_{sr}=\mu_{rd}=1$, and $\kappa_{rd}=5$, and $P/N_0\approx 20$dB  when $\eta_{sd}=\eta_{sr}=1$, $\mu_{sd}=\mu_{sr}=\mu_{rd}=2$, and $\kappa_{rd}=1$.
Secondly, there exists SNR penalty resulting with increase in the fading parameter, $\eta$, as expected. For example, an ASER of $10^{-4}$ occurs at $P/N_0\approx 24$dB  when $\eta_{sd}=\eta_{sr}=1$, $\mu_{sd}=\mu_{sr}=\mu_{rd}=1$, and $\kappa_{rd}=1$, and $P/N_0\approx 25$dB  when $\eta_{sd}=\eta_{sr}=5$, $\mu_{sd}=\mu_{sr}=\mu_{rd}=1$.
\begin{table}[t]
\begin{center}
\caption{Optimal Power Allocation Factor for
 $4\times2$-RQAM  with $\beta=1$, $\Omega_{sd}=\Omega_{sr}=1$ and $P/N_0=40\,dB$.}
\label{tableOptimal}
\begin{tabular}{|c|c|c|c|c|c|c|}
\hline
\multirow{3}{*}{$\Omega_{rd}$} & \multicolumn{3}{l|}{$\mu_{rd}=1,$}
& \multicolumn{3}{l|}{$\mu_{sd}=\mu_{sr}=1,$}   \\
 & \multicolumn{3}{l|}{$\kappa_{rd}=1$}   & \multicolumn{3}{l|}
 {$\eta_{sd}=\eta_{sr}=1$}   \\ \cline{2-7}
 &  $\mu_{sd}=\mu_{sr}$ & $\eta_{sd}=\eta_{sr}$ & $\xi$ &  $\mu_{rd}$ & $\kappa_{rd}$ &$\xi$  \\ \hline
\multirow{4}{*}{1} & \multirow{2}{*}{0.5} & 0.01 &0.7760 & \multirow{2}{*}{0.5} & 1 & 0.8000 \\ \cline{3-4} \cline{6-7}
                  &                   & 1 & 0.6544 &                   & 4 &  0.8000\\ \cline{2-7}
                  & \multirow{2}{*}{1} & 0.01 & 0.6706 & \multirow{2}{*}{1} & 1 & 0.6682 \\ \cline{3-4} \cline{6-7}
                  &                   & 1 & 0.6668 &                   & 4 & 0.6679 \\ \hline
\multirow{4}{*}{10} & \multirow{2}{*}{0.5} & 0.01 & 0.9070 & \multirow{2}{*}{0.5} & 1 &0.8000  \\ \cline{3-4} \cline{6-7}
                  &                   &  1& 0.8228 &                   & 4 & 0.8000 \\ \cline{2-7}
                  & \multirow{2}{*}{1} &0.01 & 0.6972 & \multirow{2}{*}{1} & 1 & 0.6682 \\ \cline{3-4} \cline{6-7}
                  &                   & 1 & 0.6682&                   & 4 &  0.6782\\ \hline
\end{tabular}
\end{center}
\end{table}
\subsection{Impact of Power Allocation}
Table \ref{tableOptimal} tabulates the fraction of the total power allocated to $s$ for $4\times 2$-RQAM as a function of $\Omega_{rd}$, $\eta_{sd}=\eta_{sr}$, $\mu_{sd}=\mu_{sr}$, $\kappa_{rd}$ and $\mu_{rd}$ at $P/N_0=40$dB. The values of the optimal power for any value of channel parameters of mixed $\eta$-$\mu$ and $\kappa$-$\mu$ fading channels can be obtained by solving either (\ref{Ps}) or (\ref{Pszero}) using the software MATHEMATICA. From Fig.~\ref{f6}, we can see that when fading parameters and variances of link are highly unbalance, an optimal power allocation is useful for dual-hop DF relaying system.

\section{Conclusion}
Novel closed-form expressions are obtained for the ASER of general order
RQAM and XQAM schemes with dual-hop DF relaying systems over mixed $\eta$-$\mu$ and $\kappa$-$\mu$ fading channels. The derived ASER expressions are used to analyze the performance of various QAM schemes with dual-hop DF relaying systems over mixed $\eta$-$\mu$ and $\kappa$-$\mu$ fading. We obtain the asymptotic ASER, which is useful to determine the system behavior at high SNRs in terms of the diversity order. Applying the asymptotic expression, optimal power allocation between the source node and the relay node under a total power constraint is analyzed. It is illustrated by numerical examples that the ASER performance vary according to the fading parameters of the communication links. All numerical results are validated through computer simulation results.
\begin{appendix}[Solution to the  Integrals: $\mathcal{I}_1(x,\phi)$, $\mathcal{I}_2(x,\phi)$, $\mathcal{I}_1^{\infty}(x,\phi)$ and $\mathcal{I}_2^{\infty}(x,\phi)$]
In the following subsections, we will derive the closed-form expressions for $\mathcal{I}_1(x,\phi)$, $\mathcal{I}_2(x,\phi)$, $\mathcal{I}_1^{\infty}(x,\phi)$ and $\mathcal{I}_2^{\infty}(x,\phi)$  in terms of multivariate Lauricella's $(F_D^{(n)}(\cdot), \Phi_1^{(n)}(\cdot))$ hypergeometric functions which are
given as \cite{Exton}, \cite{Martinez}
\begin{eqnarray}
\label{FD}
&&F_D^{(n)}(a,b_1,b_2,\ldots ,b_n ;c;x_1,x_2,,\ldots ,x_n)
=\frac{1}{B(a, c-a)}
\int_0^1\frac{u^{a-1}(1-u)^{c-a-1}}
{\prod_{i=1}^{n}(1-u\,x_i)^{b_i}}du
\nonumber \\ &&=\sum_{m_1,m_2=0,,\ldots,m_n=0}^{\infty}\frac{(a)_{\sum_{i}^{n}m_i}
}{(c)_{\sum_{i}^{n}m_i}}\prod_{i=1}^n\frac{(b_i)_{m_i}\,x_{i}^{m_i}}{m_i!};
|x_1|<1, |x_2|<1\ldots|x_n|<1
\end{eqnarray}
and
\begin{eqnarray}
\label{PhiD}
\Phi_1^{(n)}(a,b_1,b_2,\ldots ,b_{n-1} ;c;x_1,x_2,,\ldots ,x_n)
&&=\frac{1}{B(a, c-a)}
\int_0^1\frac{u^{a-1}(1-u)^{c-a-1}\exp(u\,x_n)}
{\prod_{i=1}^{n-1}(1-u\,x_i)^{b_i}}du
\nonumber \\ &&=\sum_{m_1,m_2=0,,\ldots,m_n=0}^{\infty}\frac{(a)_{\sum_{i}^{n}m_i \prod_{i=1}^{n-1}(b_i)_{m_i}}
}{(c)_{\sum_{i}^{n}m_i}}\prod_{i=1}^n\frac{\,x_{i}^{m_i}}{m_i!},
\end{eqnarray}
where $B(a,b)=\int_{0}^{1}x^{a-1}(1-x)^{b-1}dx$ is Beta function, $m_i!=\Gamma(m_i+1)$ and $(a)_n=\Gamma(a+n)/\Gamma(a)$ is the Pochammer symbol for $n\geq 0$. The $F_D^{(n)}\left(\cdot\right)$, and $\Phi_1^{(n)}\left(\cdot\right)$ functions can be easily and accurately evaluated by using its finite integral representation or its infinite series representation.
\subsection{Closed-form Expression for $\mathcal{I}_1\big(x,\pi/2\big)$}
Substituting (\ref{mgfsd}) into (\ref{I1}) and followed by
$u=\cos^2\theta$,
the closed-form expression for the integral $\mathcal{I}_1(x,\pi/2)$ can be given as  \cite{ErmolovaRQAM}
\begin{eqnarray}
\label{I1xpiby2}
&&\mathcal{I}_1\big(x,\pi/2\big)
=\frac{\Gamma(2\mu_{sd}+0.5)\mathcal{M}_{\gamma_{sd}}\left(x^2/2\right)}{2\sqrt{\pi}\,\Gamma(2\mu_{sd}+1)}
F_D^{(2)}\bigg(0.5,\mu_{sd},
\mu_{sd}; 2\mu_{sd}+1; \frac{4\mu_{sd}(h_{sd}-H_{sd})}{4\mu_{sd}(h_{sd}-H_{sd})+x^2\bar{\gamma}_{sd}},
\nonumber\\&&
\frac{4\mu_{sd}(h_{sd}+H_{sd})}{4\mu_{sd}(h_{sd}+H_{sd})+x^2\bar{\gamma}_{sd}}
\bigg).
\end{eqnarray}
\subsection{Closed-form Expression for $\mathcal{I}_1\big(x,\text{arccot}(y/x)\big)$}
Substituting (\ref{mgfsd}) into (\ref{I1}) and followed by
$u=1-(y^2/x^2)\tan^2\theta$,
the closed-form expression for the integral $\mathcal{I}_1(x,\text{arccot}(y/x))$ can be given as  \cite{ErmolovaRQAM}
\begin{eqnarray}
\label{Ix12}
&&\mathcal{I}_1\big(x,\text{arccot}(y/x)\big)
=\frac{x\,y\,\mathcal{M}_{\gamma_{sd}}\left(\frac{x^2+y^2}{2}\right)}{2\pi(x^2+y^2)(2\mu_{sd}+0.5)}
F_D^{(3)}\bigg(1,1,\mu_{sd},\mu_{sd};2\mu_{sd}+1.5;\frac{x^2}{x^2+y^2},
\nonumber\\&&\frac{4\mu_{sd}(h_{sd}-H_{sd})+x^2\bar{\gamma}_{sd}}{4\mu_{sd}(h_{sd}-H_{sd})+(x^2+y^2)\bar{\gamma}_{sd}},
\frac{4\mu_{sd}(h_{sd}+H_{sd})+x^2\bar{\gamma}_{sd}}{4\mu_{sd}(h_{sd}+H_{sd})+(x^2+y^2)\bar{\gamma}_{sd}}
\bigg).
\end{eqnarray}
\subsection{Closed-form Expression for $\mathcal{I}_2\big(x,\pi/2\big)$}
 Substituting, (\ref{mgfsd}) and (\ref{mgfrd}) into (\ref{I2}) followed by
$t=\frac{2\mu_{rd}^2\kappa_{rd}(1+\kappa_{rd})\sin^2\theta}
{2\mu_{rd}(1+\kappa_{rd})\sin^2\theta+x^2\bar{\gamma}_{rd}}$, and
\\ $u=\Big(\frac{2\mu_{rd}(1+\kappa_{rd})+x^2\bar{\gamma}_{rd}}{2\mu_{rd}^2\kappa_{rd}(1+\kappa_{rd})}\Big)t$, respectively,
the closed-form expression for the integral $\mathcal{I}_2(x,\pi/2)$ can be given as
\begin{eqnarray}
\label{I2apiby2}
&&\mathcal{I}_2\big(x,\pi/2\big)
=\frac{\exp(-\mu_{rd}\kappa_{rd})\Gamma(2\mu_{sd}+\mu_{rd}+0.5)x\,(2\mu_{rd}(1+\kappa_{rd}))^{\mu_{rd}}
(4\mu_{sd}\sqrt{h}(\bar{\gamma}_{rd}/\bar{\gamma}_{sd}))^{2\mu_{sd}}\sqrt{\bar{\gamma}_{rd}}}
{2\sqrt{\pi}\,\Gamma(2\mu_{sd}+\mu_{rd}+1)(x^2\bar{\gamma}_{rd}+2\mu_{rd}(1+\kappa_{rd}))^{2\mu_{sd}+\mu_{rd}+0.5}}
\nonumber\\&&\times\Phi_1^{(4)}\bigg(2\mu_{sd}+\mu_{rd}+0.5,1,\mu_{sd},\mu_{sd};2\mu_{sd}+\mu_{rd}+1;
\frac{2\mu_{rd}(1+k_{rd})}{x^2\bar{\gamma}_{rd}+2\mu_{rd}(1+k_{rd})},
\nonumber\\&&\frac{2\mu_{rd}(1+k_{rd})\bar{\gamma}_{sd}-4\mu_{sd}(h_{sd}-H_{sd})\bar{\gamma}_{rd}}
{x^2\bar{\gamma}_{sd}\,\bar{\gamma}_{rd}+2\mu_{rd}(1+k_{rd})\bar{\gamma}_{sd}},
\frac{2\mu_{rd}(1+k_{rd})\bar{\gamma}_{sd}-4\mu_{sd}(h_{sd}+H_{sd})\bar{\gamma}_{rd}}
{x^2\bar{\gamma}_{sd}\,\bar{\gamma}_{rd}+2\mu_{rd}(1+k_{rd})\bar{\gamma}_{sd}},
\nonumber\\&&
\frac{2\mu_{rd}^2\kappa_{rd}(1+\kappa_{rd})}{x^2\bar{\gamma}_{rd}+2\mu_{rd}(1+\kappa_{rd})}
\bigg).
\end{eqnarray}
\subsection{Closed-form Expression for $\mathcal{I}_2\left(x,\text{arccot}(y/x)\right)$}
Substituting, (\ref{mgfsd}) and (\ref{mgfrd}) into (\ref{I2}) followed by
$t=\frac{2\mu_{rd}^2\kappa_{rd}(1+\kappa_{rd})\sin^2\theta}
{2\mu_{rd}(1+\kappa_{rd})\sin^2\theta+x^2\bar{\gamma}_{rd}}$, and
 \\ $u=\Big(\frac{2\mu_{rd}(1+\kappa_{rd})+(x^2+y^2)\bar{\gamma}_{rd}}{2\mu_{rd}^2\kappa_{rd}(1+\kappa_{rd})}\Big)t$, respectively,
the closed-form expression for the integral $\mathcal{I}_2(x,\text{arccot}(y/x))$ can be given as
\begin{eqnarray}
\label{Ix22}
&&\mathcal{I}_2\big(x,\text{arccot}(y/x)\big)
=\frac{\exp(-\mu_{rd}\kappa_{rd})x\sqrt{\bar{\gamma}_{rd}}(2\mu_{sd}+\mu_{rd}+0.5)^{-1}(2\mu_{rd}(1+\kappa_{rd}))^{\mu_{rd}}}
{2\pi\,(4\mu_{sd}\sqrt{h}(\bar{\gamma}_{rd}/\bar{\gamma}_{sd}))^{-2\mu_{sd}}((x^2+y^2)\bar{\gamma}_{rd}+2\mu_{rd}(1+\kappa_{rd}))^{2\mu_{sd}+\mu_{rd}+0.5}}
\nonumber\\&&\times\Phi_1^{(5)}\bigg(2\mu_{sd}+\mu_{rd}+0.5,1,0.5,\mu_{sd},\mu_{sd};
2\mu_{sd}+\mu_{rd}+1.5;\frac{2\mu_{rd}(1+k_{rd})}{(x^2+y^2)\bar{\gamma}_{rd}+2\mu_{rd}(1+k_{rd})},
\nonumber\\&&\frac{x^2\bar{\gamma}_{rd}+2\mu_{rd}(1+\kappa_{rd})}{(x^2+y^2)\bar{\gamma}_{rd}+2\mu_{rd}(1+\kappa_{rd})},
\frac{2\mu_{rd}(1+k_{rd})\bar{\gamma}_{sd}-4\mu_{sd}(h_{sd}-H_{sd})\bar{\gamma}_{rd}}
{(x^2+y^2)\bar{\gamma}_{sd}\,\bar{\gamma}_{rd}+2\mu_{rd}(1+k_{rd})\bar{\gamma}_{sd}},
\nonumber\\&&\frac{2\mu_{rd}(1+k_{rd})\bar{\gamma}_{sd}-4\mu_{sd}(h_{sd}+H_{sd})\bar{\gamma}_{rd}}
{(x^2+y^2)\bar{\gamma}_{sd}\,\bar{\gamma}_{rd}+2\mu_{rd}(1+k_{rd})\bar{\gamma}_{sd}},
\frac{2\mu_{rd}^2\kappa_{rd}(1+\kappa_{rd})}{(x^2+y^2)\bar{\gamma}_{rd}+2\mu_{rd}(1+\kappa_{rd})}
\bigg).
\end{eqnarray}
\subsection{Closed-form Expression for $\mathcal{I}_1\left(x,\arctan(y/z)\right)$}
Substituting (\ref{mgfsd}) into (\ref{I1}) and followed by
$u=1-(z^2/y^2)\tan^2\theta$,
the closed-form expression for the integral $\mathcal{I}_1(x,\arctan(y/z))$ can be given as  \cite{ErmolovaRQAM}
\begin{eqnarray}
\label{Ix13}
&&\mathcal{I}_1\big(x,\arctan(y/z)\big)
=\frac{y\,z\,\mathcal{M}_{\gamma_{sd}}\left(\frac{x^2(y^2+z^2)}{2y^2}\right)}{2\pi(y^2+z^2)(2\mu_{sd}+0.5)}
F_D^{(3)}\bigg(1,1,\mu_{sd},\mu_{sd};2\mu_{sd}+1.5;\frac{y^2}{y^2+z^2},
\nonumber\\&&\frac{4\mu_{sd}(h_{sd}-H_{sd})y^2+x^2y^2\bar{\gamma}_{sd}}{4\mu_{sd}(h_{sd}-H_{sd})y^2+x^2(y^2+z^2)\bar{\gamma}_{sd}},
\frac{4\mu_{sd}(h_{sd}+H_{sd})y^2+x^2y^2\bar{\gamma}_{sd}}{4\mu_{sd}(h_{sd}+H_{sd})y^2+x^2(y^2+z^2)\bar{\gamma}_{sd}}
\bigg).
\end{eqnarray}
\subsection{Closed-form Expression for $\mathcal{I}_2\left(x,\arctan(y/z)\right)$}
Substituting, (\ref{mgfsd}) and (\ref{mgfrd}) into (\ref{I2}) followed by
$t=\frac{2\mu_{rd}^2\kappa_{rd}(1+\kappa_{rd})\sin^2\theta}
{2\mu_{rd}(1+\kappa_{rd})\sin^2\theta+x^2\bar{\gamma}_{rd}}$, and
\\ $u=\Big(\frac{2\mu_{rd}(1+\kappa_{rd})y^2+x^2(y^2+z^2)\bar{\gamma}_{rd}}{2\mu_{rd}^2\kappa_{rd}(1+\kappa_{rd})y^2}\Big)t$, respectively,
the closed-form expression for the integral \\
$\mathcal{I}_2(x,\arctan(y/z))$ can be given as
\begin{eqnarray}
\label{Ix23}
&&\mathcal{I}_2\big(x,\arctan(y/z)\big)
=
\frac{\exp(-\mu_{rd}\kappa_{rd})x y\sqrt{\bar{\gamma}_{rd}}(2\mu_{sd}+\mu_{rd}+0.5)^{-1}(2y^2\mu_{rd}(1+\kappa_{rd}))^{\mu_{rd}}}
{2\pi\,(4y^2\mu_{sd}\sqrt{h}(\bar{\gamma}_{rd}/\bar{\gamma}_{sd}))^{-2\mu_{sd}}(x^2(y^2+z^2)\bar{\gamma}_{rd}+2y^2\mu_{rd}(1+\kappa_{rd}))^{2\mu_{sd}+\mu_{rd}+0.5}}
\nonumber\\&&\times\Phi_1^{(5)}\bigg(2\mu_{sd}+\mu_{rd}+0.5,1,0.5,\mu_{sd},\mu_{sd};
2\mu_{sd}+\mu_{rd}+1.5;\frac{2\mu_{rd}(1+k_{rd})y^2}{x^2(y^2+z^2)\bar{\gamma}_{rd}+2\mu_{rd}(1+k_{rd})y^2},
\nonumber\\&&\frac{x^2y^2\bar{\gamma}_{rd}+2\mu_{rd}(1+\kappa_{rd})y^2}
{x^2(y^2+z^2)\bar{\gamma}_{rd}+2\mu_{rd}(1+\kappa_{rd})y^2},
\frac{2\mu_{rd}(1+k_{rd})y^2\bar{\gamma}_{sd}-4\mu_{sd}(h_{sd}-H_{sd})y^2\bar{\gamma}_{rd}}
{x^2(y^2+z^2)\bar{\gamma}_{sd}\,\bar{\gamma}_{rd}+2\mu_{rd}(1+k_{rd})y^2\bar{\gamma}_{sd}},
\nonumber\\&&\frac{2\mu_{rd}(1+k_{rd})y^2\bar{\gamma}_{sd}-4\mu_{sd}(h_{sd}+H_{sd})y^2\bar{\gamma}_{rd}}
{x^2(y^2+z^2)\bar{\gamma}_{sd}\,\bar{\gamma}_{rd}+2\mu_{rd}(1+k_{rd})y^2\bar{\gamma}_{sd}},
\frac{2\mu_{rd}^2\kappa_{rd}(1+\kappa_{rd})y^2}{x^2(y^2+z^2)\bar{\gamma}_{rd}+2\mu_{rd}(1+\kappa_{rd})y^2}
\bigg).
\end{eqnarray}
\subsection{Closed-form Expression for $\mathcal{I}_1^{\infty}\left(x,\pi/2\right)$}
Substituting (\ref{mgfsdapprox}) into (\ref{I1sdapprox}) and followed by
$u=\cos^2\theta$,
the closed-form expression for the integral $\mathcal{I}_1^{\infty}(x,\pi/2)$ can be given as  
\begin{eqnarray}
\label{I1approxxpiby2}
\mathcal{I}_1^{\infty}\big(x,\pi/2,\bar{\gamma}_{sd}\big)
=
\frac{B\left(0.5,2\mu_{sd}+0.5\right)}
{2\pi\,\left(4\mu_{sd}\sqrt{h_{sd}}/(x^2\,\bar{\gamma}_{sd})\right)^{-2\mu_{sd}}}.
\end{eqnarray}
\subsection{Closed-form Expression for $\mathcal{I}_2^{\infty}\big(x,\pi/2\big)$}
 Substituting, (\ref{mgfsdapprox}) and (\ref{mgfrdapprox}) into (\ref{I2srdapprox}) followed by
 $u=\cos^2\theta$, respectively,
the closed-form expression for the integral $\mathcal{I}_2^{\infty}(x,\pi/2)$ can be given as
\begin{align}
\label{I2apiby2approx}
&\mathcal{I}_2^{\infty}\big(x,\pi/2\big)
=\frac{B\left(0.5,2\mu_{sd}+\mu_{rd}+0.5\right)\exp(-\mu_{rd}\kappa_{rd})}
{2\pi\,\left(4\mu_{sd}\sqrt{h_{sd}}/(x^2\,\bar{\gamma}_{sd})\right)^{-2\mu_{sd}}}
 \left(\frac{\mu_{rd}(1+\kappa_{rd})}{x^2\bar{\gamma}_{rd}}\right)^{\mu_{rd}}.
\end{align}
\subsection{Closed-form Expression for $\mathcal{I}_1^{\infty}\big(x,\text{arccot}(y/x)\big)$}
Substituting (\ref{mgfsdapprox}) into (\ref{I1sdapprox}) and followed by
$u=1-(y^2/x^2)\tan^2\theta$, the closed-form expression for the integral $\mathcal{I}_1^{\infty}(x,\text{arccot}(y/x))$ can be given as
\begin{align}
\label{Ix12approx}
&\mathcal{I}_1^{\infty}\big(x,\text{arccot}(y/x)\big)
=\left(\frac{4\mu_{sd}\sqrt{h_{sd}}}{(x^2+y^2)\bar{\gamma}_{sd}}\right)^{2\mu_{sd}}
\frac{xy}{2\pi(x^2+y^2)}
\nonumber\\&
\times B\left(1,2\mu_{sd}+0.5\right)
F_{D}^{(1)}\left(1, 2\mu_{sd}+1;2\mu_{sd}+1.5;\frac{x^2}{x^2+y^2}\right).
\end{align}
\subsection{Closed-form Expression for $\mathcal{I}_2^{\infty}\big(x,\text{arccot}(y/x)\big)$}
Substituting, (\ref{mgfsdapprox}) and (\ref{mgfrdapprox}) into (\ref{I2srdapprox}) followed by
$u=1-(y^2/x^2)\tan^2\theta$, the closed-form expression for the integral $\mathcal{I}_2^{\infty}(x,\text{arccot}(y/x))$ can be given as
\begin{align}
\label{Ix22approx}
&\mathcal{I}_2^{\infty}\big(x,\text{arccot}(y/x)\big)
=\left(\frac{4\mu_{sd}\sqrt{h_{sd}}}{(x^2+y^2)\bar{\gamma}_{sd}}\right)^{2\mu_{sd}}
\frac{\exp(-\mu_{rd}\kappa_{rd})xy\,B\left(1,2\mu_{sd}+\mu_{rd}+0.5\right)}
{2\pi(x^2+y^2)\left((x^2+y^2)\bar{\gamma}_{rd}/(\mu_{rd}(1+\kappa_{rd}))\right)^{\mu_{rd}}}
\nonumber\\&\times
F_{D}^{(1)}\left(1, 2\mu_{sd}+\mu_{rd}+1;2\mu_{sd}+\mu_{rd}+1.5;\frac{x^2}{x^2+y^2}\right).
\end{align}
\subsection{Closed-form Expression for $\mathcal{I}_1^{\infty}\big(x,\arctan(y/z)\big)$}
Substituting (\ref{mgfsdapprox}) into (\ref{I1sdapprox}) and followed by
$u=1-(y^2/z^2)\tan^2\theta$, the closed-form expression for the integral $\mathcal{I}_1^{\infty}(x,\arctan(y/z))$ can be given as
\begin{align}
\label{Ix13approx}
&\mathcal{I}_1^{\infty}\big(x,\arctan(y/z)\big)
=\left(\frac{4y^2\mu_{sd}\sqrt{h_{sd}}}{x^2(y^2+z^2)\bar{\gamma}_{sd}}\right)^{2\mu_{sd}}
\frac{z\,y}{2\pi(y^2+z^2)}
\nonumber\\& \times B\left(1,2\mu_{sd}+0.5\right) F_{D}^{(1)}\left(1, 2\mu_{sd}+1;2\mu_{sd}+1.5;\frac{y^2}{y^2+z^2}\right).
\end{align}
\subsection{Closed-form Expression for $\mathcal{I}_2^{\infty}\big(x,\arctan(y/z)\big)$}
Substituting, (\ref{mgfsdapprox}) and (\ref{mgfrdapprox}) into (\ref{I2srdapprox}) followed by
$u=1-(z^2/y^2)\tan^2\theta$, the closed-form expression for the integral $\mathcal{I}_2^{\infty}(x,\arctan(y/z))$ can be given as
\begin{align}
\label{Ix23approx}
&\mathcal{I}_2^{\infty}\big(x,\arctan(y/z)\big)
=\left(\frac{4y^2\mu_{sd}\sqrt{h_{sd}}}{x^2(y^2+z^2)\bar{\gamma}_{sd}}\right)^{2\mu_{sd}}
\frac{\exp(-\mu_{rd}\kappa_{rd})z\,y B\left(1,2\mu_{sd}+\mu_{rd}+0.5\right)}
{2\pi(y^2+z^2)(x^2(y^2+z^2)\bar{\gamma}_{rd}/(y^2\mu_{rd}(1+\kappa_{rd})))^{\mu_{rd}}}
\nonumber\\& \times
 F_{D}^{(1)}\left(1, 2\mu_{sd}+\mu_{rd}+1;2\mu_{sd}+\mu_{rd}+1.5;\frac{y^2}{y^2+z^2}\right).
\end{align}
\end{appendix}


\end{document}